\pdfoutput=1

\documentclass[a4paper]{article}

\usepackage{amsmath,amsfonts,amssymb,graphicx,subfigure}
\usepackage[scriptsize,up,bf,sf]{caption2}

\newcommand{\vb}{$\nu_{_{b}}$}

\newcommand{\density}[2]{#1$\times$10$^{#2}\,$cm$^{-2}$}
\newcommand{\mobility}[2]{#1$\times$10$^{#2}\,$cm$^2$/Vs}

\begin{document}

\begin{center} \begin{LARGE}Observation of neutral modes in the fractional quantum Hall regime\end{LARGE}

\vspace{0.2in} \begin{large}Aveek Bid$^{1*}$, N. Ofek$^{1*}$, H. Inoue$^{1}$, M. Heiblum$^{1}$, C.L. Kane$^{2}$,\end{large}

\vspace{0.0in} \begin{large}V.Umansky$^{1}$ and D. Mahalu$^{1}$\end{large}

\vspace{0.2in} $^1$ \emph{Braun Center for Submicron Research, Department
of Condensed Matter Physics,\newline
 Weizmann Institute of Science, Rehovot
76100, Israel}

\vspace{0.0in} $^2$ \emph{Department of Physics and Astronomy, University of Pennsylvania, \newline Philadelphia,
Pennsylvania 19104, USA}

\vspace{0.0in} $^*$ \emph{These authors contributed equally to this work
}\end{center}

\begin{center}
\begin{minipage}{6in}
In the quantum Hall effect regime, taking place in a two-dimensional-electron gas under
strong magnetic field, currents flow along the edges of the sample. For some particle-hole
conjugate states of the fractional regime, e.g., with filling between 1/2 and 1 of the lowest
Landau level; early predictions suggested the presence of counter-propagating edge
currents in addition to the expected ones. When this did not agree with the measured
conductance, it was suggested that disorder and interactions will lead to counterpropagating
modes that carry only energy - the so called neutral modes. In addition, a
neutral upstream mode (Majorana mode) was also expected for selected wavefunctions
proposed for the even denominator filling 5/2. Here we report on the direct observation of
counter-propagating neutral modes in fillings 2/3, 3/5 and 5/2. This was done by injecting
such modes and allowing them to impinge on a narrow constriction, which partly reflected
them, with two main observed effects: (a) A resultant shot noise proportional to the applied
voltage on the injecting contact; (b) With simultaneously injecting also a charge mode, the
presence of the neutral mode was found to significantly affect the Fano factor and the
temperature of the backscattered charge mode. In particular, such observation for filling
5/2, may single out the non-abelian wavefunctions for the state.
\end{minipage}
\end{center}

Current propagates in the fractional quantum Hall effect (FQHE) [1] regime along the edges of a two-dimensional-electron gas (2DEG) via edge modes with a chirality dictated by the applied magnetic field [2]. While in particle-hole conjugate states, such $p-1/2<$\vb$<p$, with $p$ integer and {\vb} the Landau levels filling in the bulk, the prediction was of counter-propagating current modes [3,4], experiments did not find such edge modes [5]. Kane, Fisher and Polchinski suggested that in the presence of disorder and interactions edge reconstruction will lead to counter-propagating neutral modes, with the latter carrying only energy [6,7]. Not carrying charge these modes were difficult to detect. Since, to the best of our knowledge, the neutral modes have not been observed thus far [8] (hence, sometimes called `elusive'), it is not surprising that very little is known about them. For example, unknown is the energy they carry; their interactions with potential barriers; their decay length; their temperature dependence; their velocity; and their interaction with charge modes.

Proposals of how to detect the neutral modes involve measuring tunneling exponents in constrictions [7]; observing thermal transport [9,10]; searching for resonances in a long constriction [11]; or looking for heating effects on charge modes [12,13]. Alternatively, our approach was to allow an upstream neutral mode, if it were to exist, to impinge on a quantum point contact (QPC) constriction with the hope that the neutral quasiparticles will be fragmented into charge carriers. Since the neutral mode is not expected to carry average current, the fragmentation was tested via measuring shot noise. By injecting simultaneously also a charge mode, we could also measure the effect of the neutral mode on the transmission probability $t$ of
the QPC constriction and on the shot noise of the partitioned charge mode. While $t$ was found to depend very weakly, the shot noise due to the charge mode was found to be highly sensitive to the presence of the neutral mode. We studied in some detail the `model' on the fractional state $\nu_b=2/3$ and present also data, albeit more briefly, for \vb=3/5, \vb=5/3 and \vb=5/2. For comparison, similar experiments were also done for `regular' states \vb=1, \vb=2/5 and \vb=1/3 - proving the absence of such striking effects. We stress that we concentrate here mainly on the observation of neutral modes and not on many of their unique properties, which are now under investigation.

\begin{center}
\textbf{Neutral edge in the \vb=2/3 state}
\end{center}

At \vb=2/3 in an ideal 2DEG, with a rather fast charge density drop toward the edge of the sample, it was predicted that two spatially separated edge modes coexist: an electron channel moving downstream close to the sample's edge and an inward $e^2/3h$ ($e$ electron charge and $h$ the Planck constant) upstream channel [3,4]. This picture can also be explained with the composite fermion (CF) model [14] - applicable to fractional states in the lowest Landau level. This two channel model predicts a two-terminal conductance of $(4/3)e^2/h$, which was never observed. When electron interactions and disorder are taken into account, mixing of the two oppositely propagating charge modes is expected to result with a downstream mode of conductance $(2/3)e^2/h$ and an upstream neutral mode [6,7,9]; agreeing with the measured two-terminal conductance $(2/3)e^2/h$. One can view the neutral mode as a fluctuating `dipole' that propagates at a lower velocity than the charge mode velocity [7,11] (or even at zero velocity, [15,16]); decaying with distance and with temperature as $T^{-2}$ [7].

\begin{center}
\textbf{Sample and setup}
\end{center}

The configuration of the sample (used for all filling factors except for the \vb=5/2), fabricated in a GaAs-AlGaAs heterostructure with an embedded 2DEG, is shown in Fig. 1. The 2DEG, with carrier density \density{1}{11} and dark mobility $>$\mobility{10}{6} at $T<$1K, was buried 116nm below the surface of the heterostructure. A $\sim$100nm long negatively biased split-gate (15nm Ti / 30nm Au) with an opening $\sim$600nm wide formed the QPC constriction. The grounded contacts (made of AuGeNi) were tied directly to the \emph{cold finger} of the dilution refrigerator at $\sim$10mK, thus cooling effectively the electrons to $\sim$10mK (verified by noise measurements). The magnetic field was raised to B=6.4Tesla leading to \vb=2/3 - as identified by $R_{xx}\sim0$ (at the bulk and also through the QPC) and a Hall plateau $R_{xy}\cong39k\Omega$. Current was injected from source \#1 ($I_s$) with a counter clockwise chirality, directing the current towards the QPC constriction (transmission probability $t$). Generated shot noise was collected by the voltage probe (with a LC resonant circuit tuned to 770kHz with bandwidth $\sim$40kHz). The signal was first amplified by a cooled home-made preamplifier (voltage gain $\sim$7), which was followed by a room temperature amplifier (NF-220F5) with voltage gain $\sim$200 and a spectrum analyzer. From the opposite side of the mesa current was injected from source \#2 ($I_n$), propagating downstream away from the QPC constriction and collected by ground \#1. A neutral mode, if it were to exist, was expected to emanate from source \#2 and move upstream towards the QPC constriction, which was $\sim$40$\mu$m (or $\sim$40$\mu$m away). Similarly, source \#3 could be charged too.

\begin{figure}[!h]
\begin{center}
  \includegraphics*[width=5in]{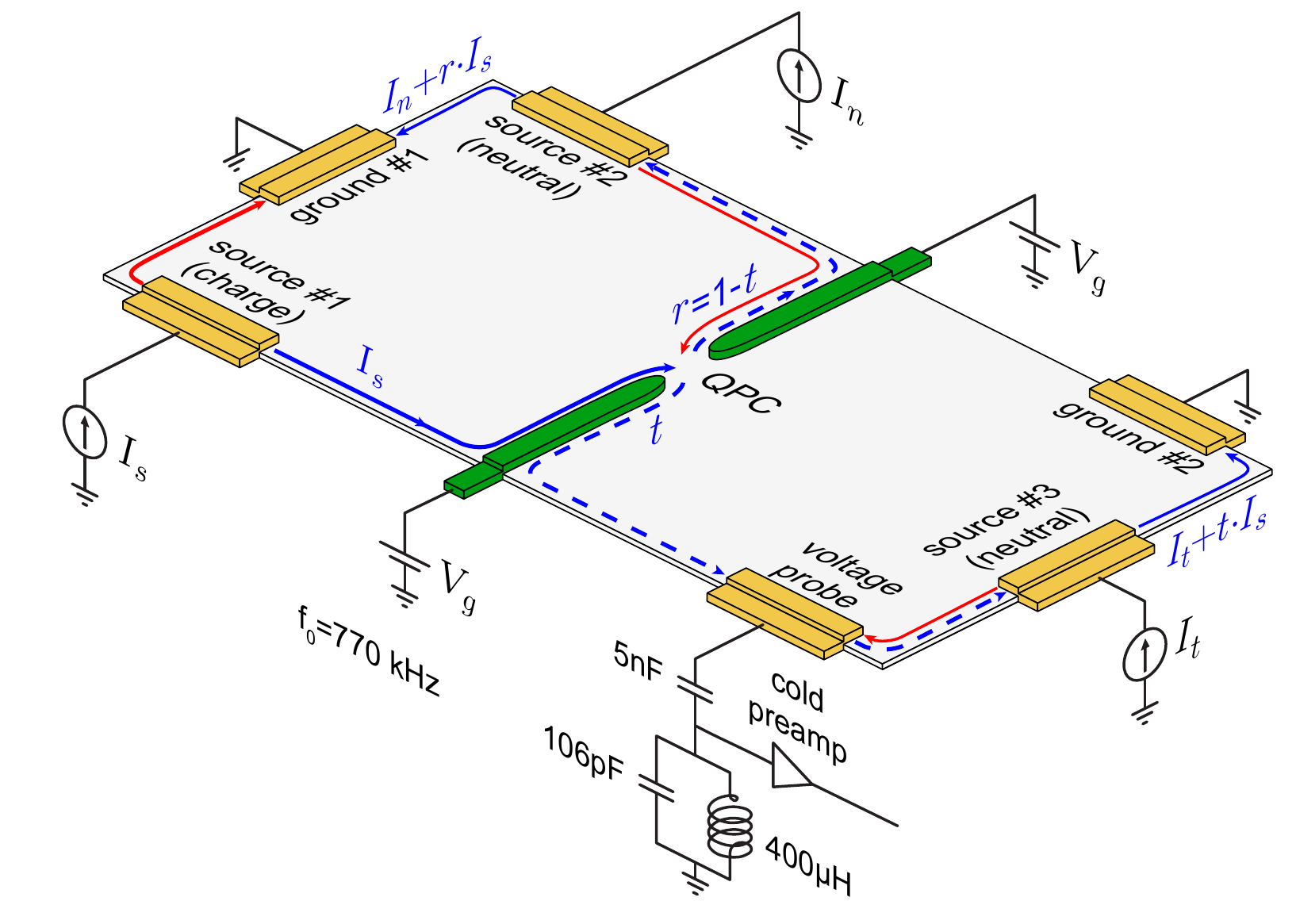}
\begin{minipage}{6in}
  \caption{The experimental setup for measuring the neutral mode. The orange pads are ohmic contacts. The green pads form a split-gate of the QPC constriction, with $V_g$ controlling the transmission probability $t$. The grounded contacts are directly connected to the cold finger of the dilution refrigerator. Excitation current is driven to the sources via a DC voltage V and a large resistor in series (1G$\Omega$). A small AC signal is used to measure the two-terminal differential conductance. Blue lines describe the downstream charge edge modes, while red lines stand for the upstream neutral edge modes. Note that due to the multi-terminal configuration the `current noise' of the preamplifier (injected backwards from the preamplifier's input into the sample) and the measured thermal noise were both independent of $t$ [17]. The cryogenics preamplifier's current noise was $\sim$13.6fA/$\sqrt{\mbox{Hz}}$ and its voltage noisewas 680pV/$\sqrt{\mbox{Hz}}$; both referred to its input.}
  \end{minipage}
\end{center}
\end{figure}

\begin{center}
\textbf{Shot noise}
\end{center}

What is the expected noise at the voltage probe? The total noise is composed of shot noise that exists only when current is driven (termed `excess noise'); thermal (Johnson-Nyquist) noise; background noise, mainly due to 1/$f$ noise ($f$ the frequency) being negligible at the frequency of measurement, and instrumentation noise. For stochastic backscattering events by the QPC constriction, injecting a noiseless current from a source at zero temperature is expected to lead to a binomial charge distribution in the partitioned current [17-21]. For single edge channel transport, partitioning of $e^*$ charges at finite temperature was found to be also stochastic under certain conditions, with low frequency spectral density of the excess noise and thermal noise, $S^i\left(V_s,\omega\sim0\right)_T$ [17]:
\begin{equation}
  S^i\left(V_s,0\right)_T=2e^*V_sg_{_{b}}t(1-t)\left[\coth\left(\frac{e^*V_s}{2k_{_{B}}T}\right)-\frac{2k_{_{B}}T}{e^*V_s}\right]+4k_{_{B}}Tg_{_{b}},
\end{equation}
where $V_s$ is the applied source DC voltage, $g_{_{b}}=$\vb$\,e^2/h$ is the Hall conductance. Empirically,
all the noise measurements here complied with this form. The dependence of the excess shot
noise is captured by the inferred quasiparticles temperate ($T$) and its effective charge ($e^*$). We
will describe the effect of $I_n$ on the noise throughout these two parameters. Note that lower lying
channels, which traverse the constriction with unity transmission probability, do not carry excess
noise [21].

\begin{center}
\textbf{Measurements at \vb=2/3 state}
\end{center}

Sources (\#1, \#2, \#3 in Fig. 1) were charged separately: (a) Charging source \#3, hence injecting $I_t$
counter clockwise and a neutral mode clockwise towards the voltage probe - did not add any
measurable noise at the voltage probe (our temperature resolution was $<$2mK/$\sqrt{\mbox{Hz}}$); (b)
Reversing the polarity of the magnetic field and then charging source \#1, thus injecting $I_s$
clockwise and a neutral mode toward the QPC constriction, led again to a null added noise,
independent of $t$; and (c) Back in the original orientation of the magnetic field, charging source
\#2, thus injecting $I_n$ counter clockwise and a neutral mode clockwise, led to a significant excess
noise in the voltage probe for $t<1$ (see Fig. 2). The excess noise, which increased initially almost
linearly with $I_n$ tended to saturate for $I_n>2$nA. Moreover, it was seemingly proportional to $t(1-t)$
(with zero excess noise when $t=1$ and $t=0$ and with a maximum at $t\sim$1/2).

\begin{figure}[!h]
\begin{center}
  \includegraphics*[width=5in]{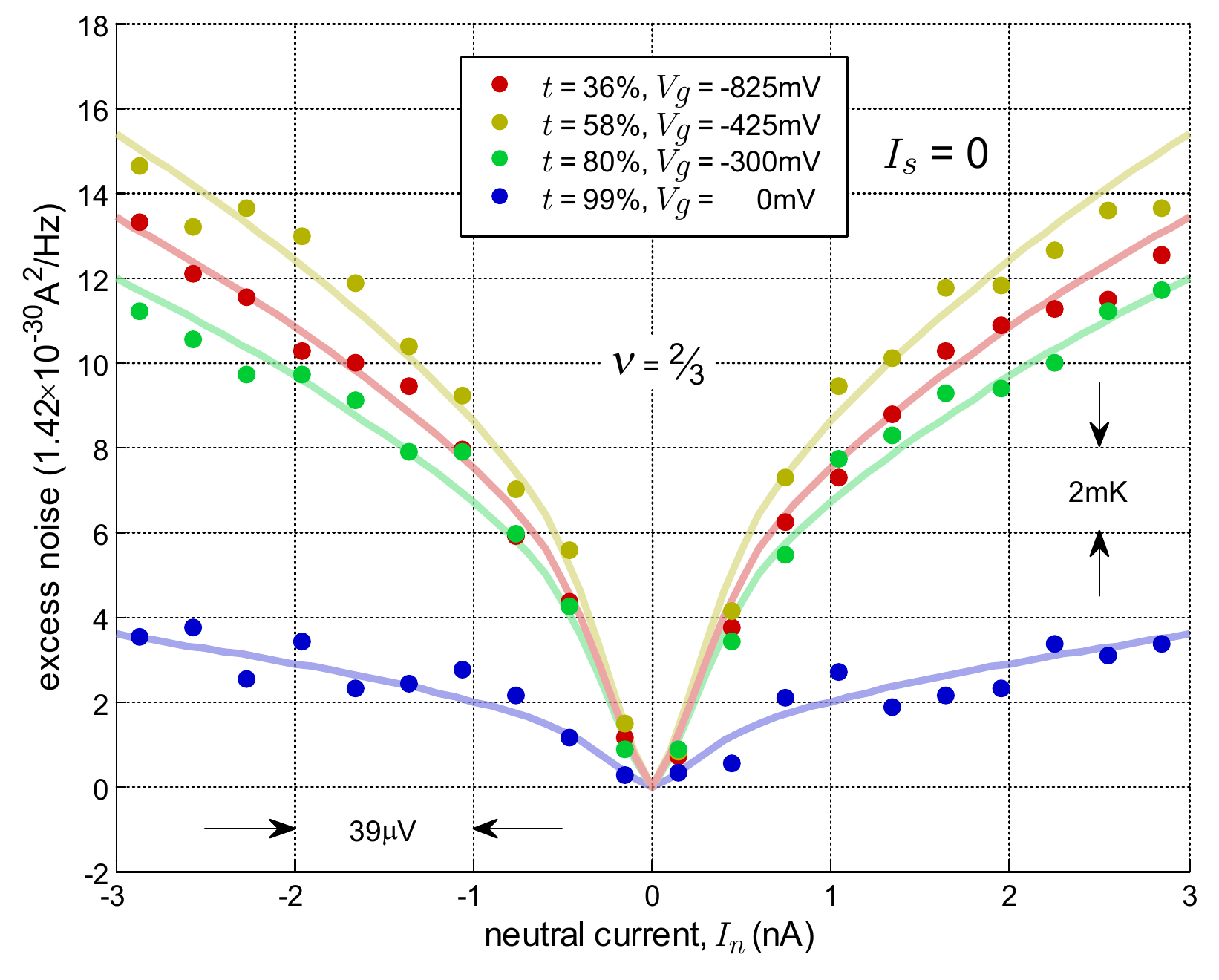}
\begin{minipage}{6in}
  \caption{Detection of the neutral mode at \vb=2/3. Excess noise measured at the voltage probe as a function of driven current $I_n$ via source \#2, for different transmission probabilities $t$ of the QPC constriction. The noise is proportional to $I_n$ and approximately to $t(1-t)$ - vanishing for $t=1$ and $t=0$. The voltage bar is $\Delta V_n=39\mu$V for $\Delta I_n=1$nA and the temperature bar $\Delta T$=2mK is calculated via $S^i=4k_{_{B}}g_{_{2/3}}\Delta T$.}
  \end{minipage}
\end{center}
\end{figure}

These results can be understood qualitatively if indeed an upstream neutral mode exists. When source \#2 is charged, some of the power dissipated at source \#2 can excite the upstream neutral mode there. When incident on the QPC, the excited neutral mode leads to enhanced fluctuations in the charge crossing the QPC. This can be modeled as if neutral quasiparticles (`dipoles') were fragmented into partitioned quasiparticles and quasiholes, or in a manner similar to Johnson noise, which occurs when thermal energy is present in all the incident channels. The current noise generated at the QPC then follows the chirality of the charge mode and is detected in the voltage probe (note that the chirality of electrons and holes in the conduction band is similar). Thus, the QPC effectively converts the upstream neutral current into a measurable charge noise signal. To get an order of magnitude, the excess noise for $I_n$=2nA is equivalent to shot noise generated by $\sim$250pA carried by electrons.

We test now the interaction of the neutral mode with a charge mode at the QPC constriction. Excess noise of partitioned charge modes had been already measured at \vb=2/3 [22]. The backscattered quasiparticle charge was found to be strongly temperature dependent, with $e^*$=(2/3)$e$ at $T${$\sim$}10mK over a wide range of $t$, dropping to $e^*${$\sim$}{$e$}/3 around $T${$\sim$}120mK (the charge evolution is shown again for convenience in Fig. 3d). Injecting $I_s$ (from source \#1 while $I_n$=0) led, again, to an excess noise and $e^*${$\sim$}(2/3)$e$ for $t$=0.3$-$0.8. Measurements of the non linear transmission and shot noise were repeated when source \#2 was also charged, thus injecting a neutral mode toward the QPC constriction (Fig. 3a). While the transmission changed merely by a fraction of a percent, the noise was affected dramatically by $I_n$. These results can be understood in the following way: (a) Charging source \#2 added a constant noise at the voltage probe (seen for $I_s$=0 in Fig 3a, and being symmetric with respect to $\pm V_n$ ); (b) The partitioned quasiparticle charge dropped down to $e^*${$\sim$}0.4$e$ at $I_n$=3nA ($\Delta V_n${$\sim$}120$\mu$V) (Fig. 3b); and (c) The apparent temperature of the partitioned quasiparticles increased by $\Delta T_{qp}${$\cong$}15mK at $I_n$=-2nA - with temperature reaching 25-30mK (Fig. 3c) - as determined from the `increased rounding' in the spectral density near $I_s$=0 (see Eq. 1). This relatively small temperature increase cannot account for the charge evolution shown in Fig. 3b (see temperature dependence in Fig. 3d).

\begin{figure}[!h]
\begin{center}
  \includegraphics*[width=6in]{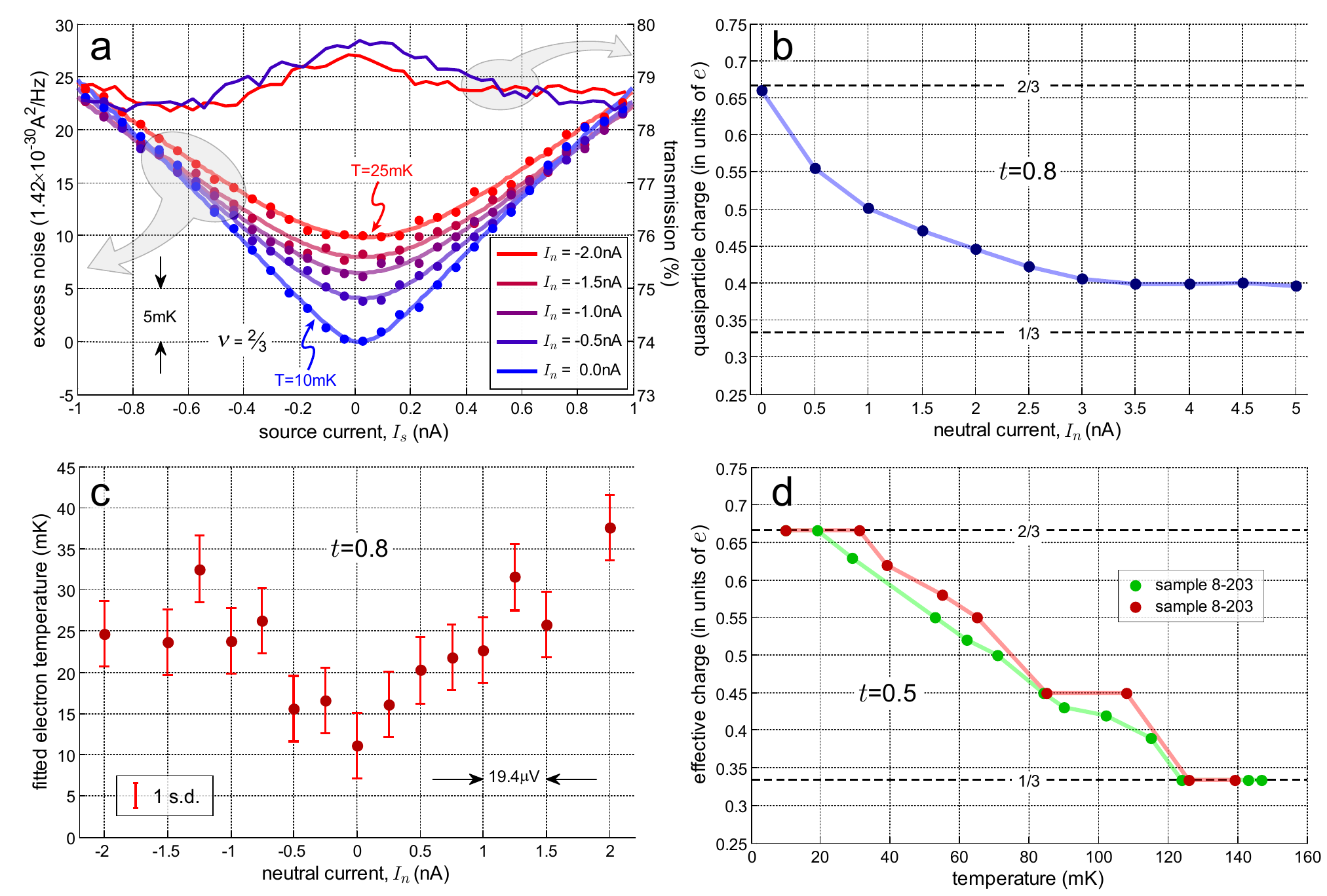}
\begin{minipage}{6in}
  \caption{The effect of impinging the neutral mode simultaneously with the charge mode on the QPC constriction at \vb=2/3. (a) The conductance and total noise as a function of source \#1 current $I_s$ for different values if $I_n$. The change in the non-linear transmission probability as $I_n$ increases is negligible. The excess noise increases, the partitioned quasiparticle charge diminishes, and the temperature of the quasiparticles increases as $I_n$ increases. (b) Charge evolution as a function of $I_n$. The charge starts at $e^*$=(2/3)$e$ and drops to $e^*$=0.4e. (c) Temperature evolution of the partitioned quasiparticles as a function of $I_n$. The temperature, fitted from (a), increases by 15-25mK at $I_n$=2nA. (d) The dependence of the quasiparticle charge as a function of temperature (see Ref. 22).}
  \end{minipage}
\end{center}
\end{figure}

Similar measurements were performed at different fridge temperatures and also using a QPC located $\sim$120$\mu$m from the source. From figure 4a it is clear that the excess shot noise due to $I_n$ diminishes with temperature. Increasing the distance by 80$\mu$m also reduce the excess shot noise, but differently at different temperatures. At T~10mK, these extra 80$\mu$m cause a reduction of $\sim$10\% and at $T${$\sim$}25mK is cause a reduction of $\sim$40\%. These corresponds to $x_0=380\mu$m and $x_0=80\mu$m respectively. It indeed scales approximately as $T^{-2}$.

\begin{figure}[!h]
\begin{center}
  \includegraphics*[width=6in]{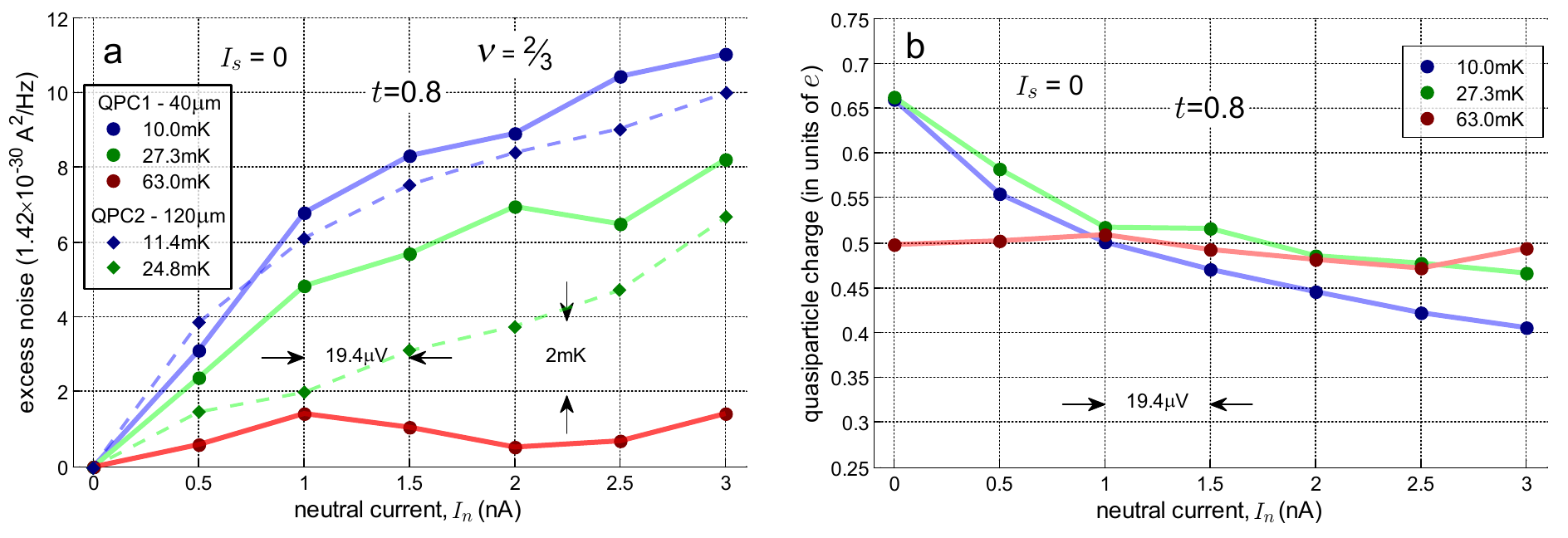}
\begin{minipage}{6in}
  \caption{The effect of neutral edge as a function of temperature and distance from the neutral source at \vb=2/3 (a) Excess noise at $\nu=2/3$ as a function of $I_n$ for different temperatures and two distances between the Source \#2 and the QPC. (b) Charge evolution as a function of $I_n$ at different base temperatures.}
  \end{minipage}
\end{center}
\end{figure}

We move to see how the temperature affects the charge dependence on $I_n$. For these measurements we stabilize the fridge each time at a different base temperature. We then measure shot noise as a function of $I_s$, at fixed values of $I_n$ and extract the charge. The QPC is set to the same transmission in all the measurements. As already pointed out, applying $I_n$ lowers the effective charge. This effect is more dramatic at low temperature ($\sim$10mK) and becomes milder with temperature, with almost no effect on the backscattered quasiparticle charge at $T$=63mK. Note that as the temperature was increased, the quasiparticle charge at $I_n$=0 was lower than (2/3)$e$ - corroborating the results of Fig. 3d.

Before presenting results for three more fractional states that were theorized to posses an upstream neutral mode, \vb=3/5, \vb=5/3 and \vb=5/2, we bring evidence that `simpler' fractional states, such as \vb=1/3, \vb=2/5 and \vb=1, do not support upstream neutral modes (in general the states are with $p<v<p+1/2$, with $p$ zero or an integer). We start with \vb=2/5 because its partitioned fractional charge was found also to evolve with temperature in a fashion similar to that of \vb=2/3, namely, the weakly backscattered quasiparticle charge was $e^*$=(2/5)$e$ at 10mK, dropping to $e^*$=$e$/5 at approximately 50mK [23] (hence, no change in the noise will exclude a simple `heating' effect caused by $I_n$). Increasing the field to B=10.5Tesla (corresponding to \vb=2/5), we first charged source \#2 with $V_s$=0; without an observed increase in the excess noise for two different transmissions (Fig. 5a). Performing conductance and noise measurements as function of $I_s$, at different values of $I_n$ ($I_n$=0$-$3nA), did not show, again, any effect of $I_n$ (Fig. 5b). These results are in overwhelming contrast with those at \vb=2/3, excluding the presence of an upstream neutral mode. Similar measurements were performed at \vb=1/3 and \vb=1, and again, no measurable effects were observed when the neutral contact (source \#2) was charged (we didn't find it necessary to show the null results again).

\begin{figure}[!h]
\begin{center}
  \includegraphics*[width=6in]{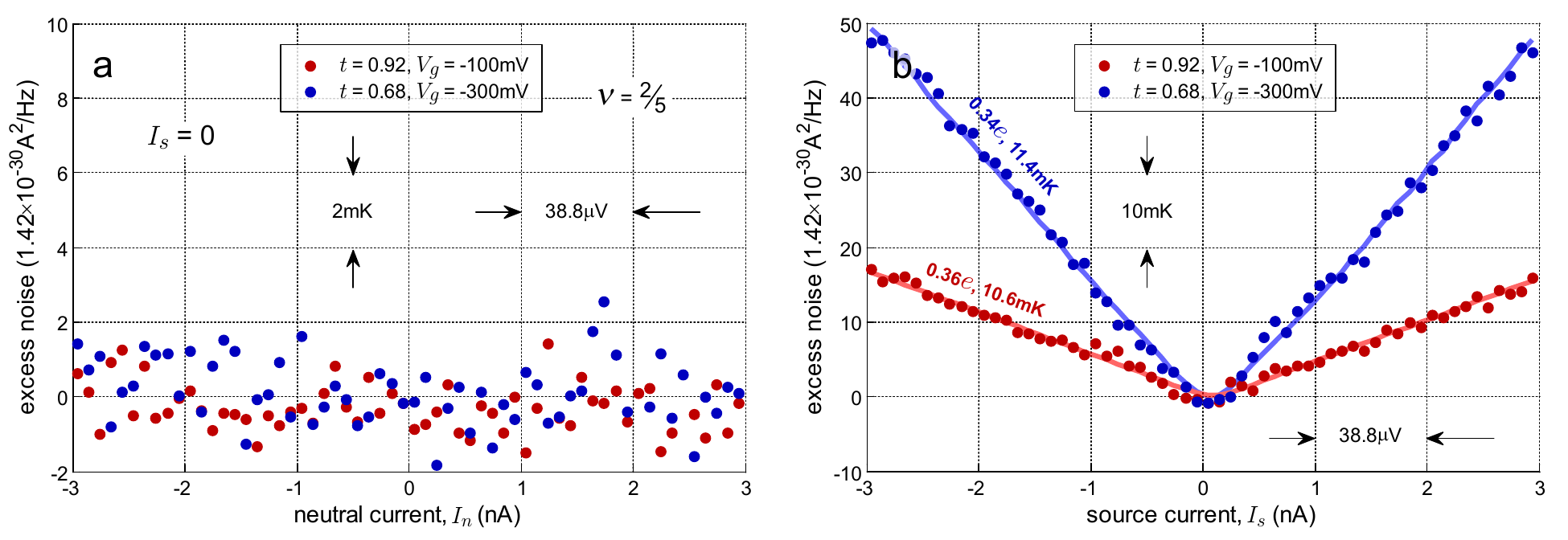}
\begin{minipage}{6in}
  \caption{Measurements at fractional state \vb=2/5. (a) Injecting only $I_n$, with two different transmissions of he QPC constriction, did not result in any excess shot noise. (b) Similarly, injecting both, $I_n$ and $I_s$ and plotting the excess noise as a function of $I_s$ at two different transmissions did not have any effect on the excess noise.}
  \end{minipage}
\end{center}
\end{figure}

\begin{center}
\textbf{Measurements at \vb=3/5 state}
\end{center}

We continue with the fractional state \vb=3/5. Being the particle-hole conjugate of \vb=2/5, it is expected to support two upstream neutral modes and one downstream charge mode [7,9]. Tuning the field to B=7Tesla with a clear fractional state \vb=3/5, charging source \#3 did not lead to any increase in the noise at the voltage probe. However, as for \vb=2/3, injecting the neutral mode by charging source \#2 with $t<$1 of the QPC constriction, led to excess noise nearly
\begin{figure}[!h]
\begin{center}
  \includegraphics*[width=6in]{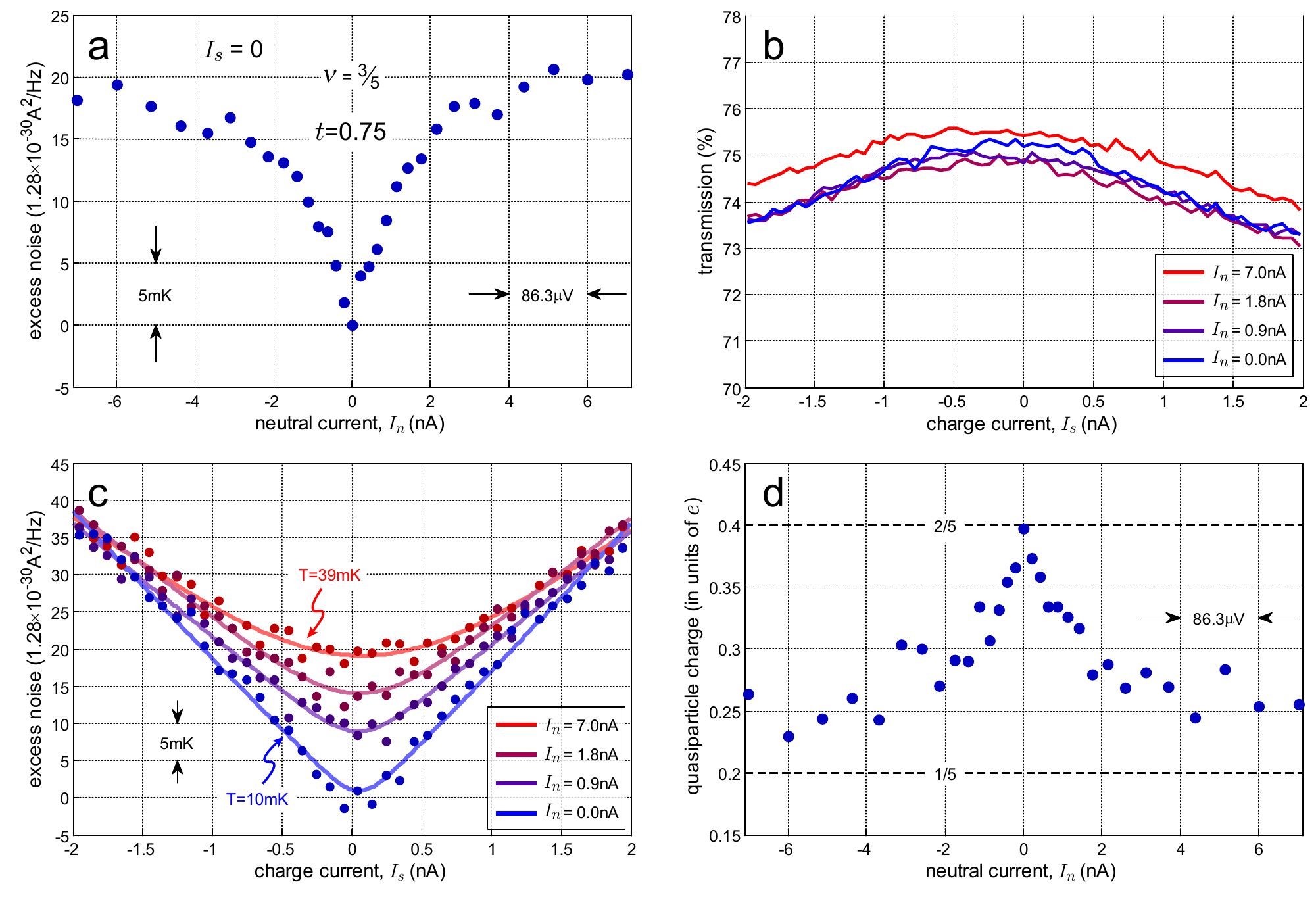}
\begin{minipage}{6in}
  \caption{Testing for the existence of the neutral mode at \vb=3/5. (a) Excess noise as a function of the injected $I_n$ - direct evidence of an upstream neutral mode. (b) The dependence of the nonlinear conductance as a function of $I_s$ on the presence of $I_n$. The relative change is small, mounting to a fraction of a percent. (c) The dependence of the excess noise as a function of $I_s$ on the presence of $I_n$. The noise increases, the quasiparticle charge drops, and the temperature increases. (d) The dependence of the quasiparticle charge on $I_n$ [extracted from (c)].}
  \end{minipage}
\end{center}
\end{figure}

linear with $I_n${$<$}1nA, and tending to saturate for higher values of $I_n$ (Fig. 6a). The excess noise (or, the equivalent temperature) was more than 50\% higher than in \vb=2/3; possibly accounting for the two upstream neutral modes in the \vb=3/5 state. Charging source \#1 in the presence of charged source \#2, the presence of $I_n$ affected only slightly the non-linear conductance (by a fraction of a percent, Fig. 6b); however, again, the altered the excess noise significantly (Fig. 5c). As before, the determined  backscattered quasiparticles charge dropped with $I_n$ from $e^*${$\sim$}(2/5)$e$ at $I_n$=0 to $e^*$=0.25$e$ at $I_n${$\sim$}5nA (Fig. 6c and 6d). As evident in Fig. 5c, the temperature of the partitioned quasiparticles increased as $I_n$ increased.

We also tested the \vb=5/3 (\vb=1+2/3) fractional state. Unlike in the \vb=2/3 state, this state is expected to support two downstream modes and only one upstream mode. Unfortunately, we are not aware of a theoretical treatment of this complex edge mode with interactions and in the presence of disorder. In the measurements we have observed a barely marginal effect of an upstream neutral mode, leaving this fractional state for future studies.

\begin{center}
\textbf{Measurements at \vb=5/2 state}
\end{center}

We turn now to \vb=5/2 state. While the expectations are that this fractional state is of nonabelian nature, thus supporting a neutral Majorana mode, the nature of the state as well as the presence of the neutral mode were not established thus far. A Pfaffian state with an unreconstructed edge will not have an upstream neutral mode [24]. An anti-Pfaffian state with a disorder-dominated but unreconstructed edge will have three upstream neutral Majorana modes [10,25]. If the edge is reconstructed, as may be expected for a smooth confining potential, then the Pfaffian and anti-Pfaffian states can both have a single upstream neutral Majorana mode [26]. An experiment which could distinguish these four possibilities has been proposed in Ref. 12. If the edges were not disorder-dominated, then the anti-Pfaffian state would have the wrong two-terminal conductance for the same reason as the \vb=2/3 state [10,25]. However, in the absence of a microscopic theory it is very difficult to make definite statements, and thus a detection of an upstream neutral mode can only strengthen the belief of the non-abelian nature of the \vb=5/2 state.

For these measurements a different heterostructure was used with the same contacts configuration. The details of such a heterostructure were already reported before [27] (see also Fig. 7 caption). Clear signatures of the \vb=5/2 state were observed with $R_{xx}${$\sim$}0 at B=5Tesla. The first and the most important result is shown in Fig. 7a, where only source \#2 was charged. Excess noise was observed with an approximate quadratic increase with $I_n$. This proves, right from the start, the presence of an upstream neutral mode. While the increase of the noise was the smallest among the fractional state being tested, it was in relative terms the highest since the actual current that was carried by the fractional state was the smallest (as 4/5 of the current is carried by the first two, lower lying, integer Landau levels). Similarly, the excess noise due to current arriving from source \#1, when charged, was strongly affected by $I_n$; with an apparent increase of the quasiparticles temperature while the quasiparticle charge dropped (Fig. 7c and 7d). Again, like in the \vb=2/3 state, this temperature increase cannot account for the charge drop [28]. The charge dropped with In from $e^*$=0.75$e$ to $e^*$=0.32$e$ at $I_n$=10nA. A similar evolution of the charge, but as a function of temperature, had been reported recently [28]. The non-linear transmission also changed, albeit by a very small amount (Fig. 7b).

\begin{figure}[!h]
\begin{center}
  \includegraphics*[width=6in]{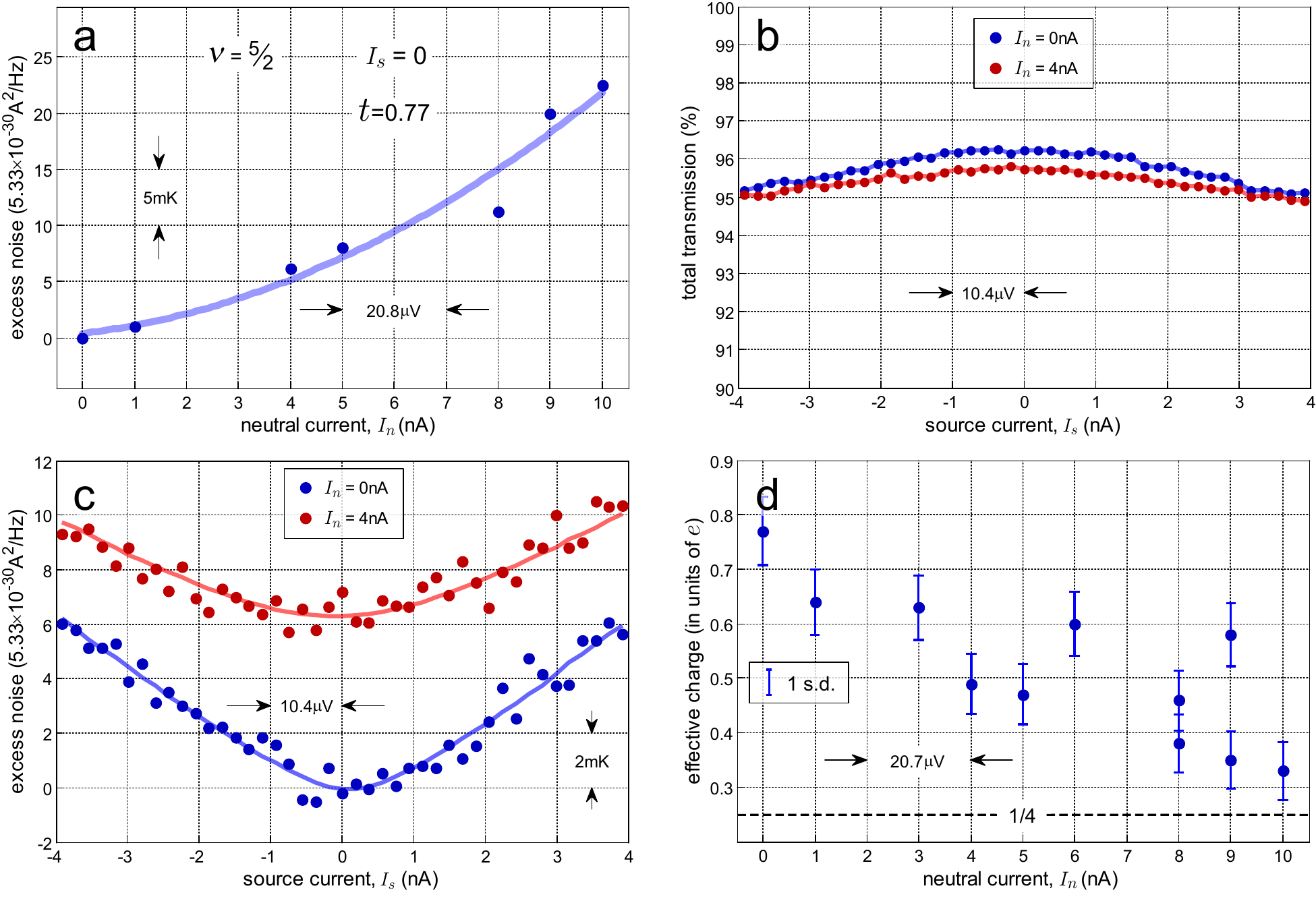}
\begin{minipage}{6in}
  \caption{Testing for the existence of the neutral mode at \vb=5/2. The 2DEG used for these measurements was embedded in a 30nm wide quantum well, which was doped on both sides, buried approximately 160nm below the surface of the heterostructure. The carrier density was \density{3.10}{11} and the low temperature dark mobility was $>$\mobility{30}{6}. (a) Excess noise as a function of injecting In provides direct evidence of an upstream neutral mode. (b) The dependence of the non-linear conductance as a function of $I_s$ on the presence of $I_n$. The relative change in the transmission is very small, mounting to much less than 1\%. Since 80\% of the total current flows in the two underlying edge channels (\vb=1 and \vb=2), the effective transmission is about 77\%. (c) The dependence of the excess noise as a function of $I_s$ on the presence of $I_n$. (d) The dependence of the quasiparticle charge on $I_n$ [extracted from (c)]. The charge drops from $e^*${$\sim$}0.75$e$ to $e^*${$\sim$}0.32$e$.}
  \end{minipage}
\end{center}
\end{figure}

\begin{center}
\textbf{Discussion}
\end{center}

Presenting the first evidence of the existence of neutral modes in the fractional states \vb=2/3, \vb=3/5 and \vb=5/2, using a ubiquitous QPC constriction serving as a detector, the following findings can be summarized: (i) A flux of neutral quasiparticles, emitted from a biased ohmic contact does not carry current or shot noise. Moreover, a neutral mode impinging on a macroscopic ohmic contact, does not increase its temperature by a measurable amount. (ii) An impinging flux of neutral quasiparticles on a QPC constriction having a finite transmission $t$, results in excess shot noise. The excess noise is approximately proportional to $t(1-t)$ and to the voltage of the injecting contact. The upstream energy flux, in the odd denominator fractions, seems to be correlated with the ratio between the number of upstream and downstream modes. (iii) Having a neutral mode impinging on a QPC constriction, while a charge mode is simultaneously being partitioned, alters dramatically the noise and the deduced partitioned quasiparticle charge. The charge drops inversely in proportion to the injecting voltage. (iv) In the same experiment, the temperature of the simultaneously partitioned quasiparticles increases with increasing the injecting voltage. However, the temperature increase is too small to account for the observed drop in charge. The mechanism responsible for modifying the tunneling cross-section of the quasiparticles in the QPC constriction is not currently understood. (v) Assuming a temperature dependent energy decay of $T^{-2}$, the typical length scale is $\sim$100$\mu$m at 25mK for \vb=2/3. (vi) Observing an upstream neutral mode in the even denominator fraction \vb=5/2 rules out, according to present theories, an abelian wavefunction of this state, and thus narrows down the spectrum of possible states (see above).

We trust that with this relatively easy method of observing the so called `elusive neutral modes', new studies of their properties will be launched, possibly shedding new light on the characteristics of edge mode transport in fractional states – not revealed via their charge carrying nature.

\begin{center}
\textbf{Acknowledgments}
\end{center}

We thank D. Feldman, Y. Gefen, A. Stern, Y. Gross and I. Neder for valuable discussions and E.
Alcobi for comments on the manuscript. The work enjoyed partial support of the Israeli Science
Foundation (ISF), the Minerva foundation, the German Israeli Foundation (GIF), the German
Israeli Project Cooperation (DIP), the European Research Council under the European
Community's Seventh Framework Program (FP7/2007-2013)/ERC Grant agreement \# 227716,
and the US-Israel Bi-National Science Foundation. N. Ofek acknowledges support from the
Israeli Ministry of Science and Technology. C. Kane acknowledges support from NSF Grant
DMR 0906175.

\begin{center}
\textbf{References}
\end{center}

\scriptsize{\begin{enumerate}
\item Das Sarma, S. \& Pinczuk, A., "Perspective in Quantum Hall effects: Novel Quantum Liquid in Low-Dimensional Semiconductor Structures". (New York, Wiley), 1997.
\item Wen, X. G., "Chiral Luttinger liquid and the edge excitations in the fractional quantum Hall states". Phys. Rev. B \textbf{41}, 12838-12844 (1990).
\item MacDonald, A. H. "Edge states in fractional quantum Hall effect regime". Phys. Rev. Lett. \textbf{64}, 220-223 (1990).
\item Johnson, M. D. \& MacDonald, A. H. "Composite Edges in v=2/3 fractional quantum Hall
effect". Phys. Rev. Lett. \textbf{67}, 2060-2063 (1991).
\item Ashoori, R. C., Stormer, H. L., Pfeiffer, L. N., Baldwin, K. W. \& West, K. "Edge
magnetoplasmons in time domain". Phys. Rev. B \textbf{45}, 3894-3897 (1992).
\item Kane, C. L., Fisher M. P. A. \& Polchinski. J. "Randomness at the edge: theory of quantum Hall transport at filling v=2/3". Phys. Rev. Lett. \textbf{72}, 4129-4132 (1994).
\item Kane, C. L. \& Fisher M. P. A. "Impurity scattering and transport of fractional quantum Hall edge states". Phys. Rev. B \textbf{51}, 13449-13466 (1995).
\item Granger, G., Eisenstein, J. P. \& Reno, J. L. "Observation of chiral heat transport in the quantum Hall regime". Phys. Rev. Lett. \textbf{102}, 086803 (2009).
\item Kane, C. L. \& Fisher M. P. A. "Quantized thermal transport in the fractional quantum Hall effect". Phys. Rev. B. \textbf{55}, 15832-15837 (1997).
\item  Levin, M., Halperin, B. I. \& Rosenow. B. "Particle-hole symmetry and the Pfaffian state". Phys. Rev. Lett. \textbf{99}, 236806 (2007).
\item Overbosch, B. J \& Chamon, C. "Long tunnelling contact as a probe of fractional quantum Hall neutral edge modes". Phys. Rev. B \textbf{80}, 035319-035324 (2009).
\item D. E. Feldman, D. E. \& Li, F. "Charge-statistics separation and probing non-Abelian states". Phys. Rev. B \textbf{78}, 161304-161307 (2008).
\item Grosfeld, E. \& Das, S. "Probing the neutral edge modes in transport across a point contact via thermal effects in the Read-Rezayi non-abelian quantum Hall states". Phys. Rev. Lett.  \textbf{102}, 106403 (2009).
\item Johnson, B. L. \& Kirczenow, G. "Composite fermions in quantum Hall effect". Rep. Prog. Phys. \textbf{60}, 889-939 (1997).
\item Lopez, A. \& Fradkin, E. "Universal structure of the edge states of the fractional quantum Hall states". Phys. Rev. B \textbf{59}, 15323-15331 (1999).
\item Lee, D. H. \& Wen, X. G. "Edge tunneling in fractional quantum Hall regime". Cond-Mat
9809160v2 (1998).
\item Heiblum, M. "Quantum Shot Noise in Edge Channels". Phys. Stat. Sol. (b) \textbf{243}, 3604-3616 (2006).
\item Lesovik, G. B. "Excess quantum shot noise in 2D ballistic point contacts". JETP Letters \textbf{49}, 592-594 (1989).
\item Kane, C. L. \& Fisher, M. P. A. "Nonequilibrium noise and fractinal charge in the quantum Hall effect". Phys. Rev. Lett. \textbf{72}, 724-727 (1994).
\item Martin, T. \& Landauer, R. "Wave packet approach to noise in multichannel mesoscopic
systems". Phys. Rev. B \textbf{45}, 1742-1755 (1992).
\item Dolev, M, Heiblum, M., Umansky, V., Stern, Ady, \& Mahalu, D. "Observation of a quarter of electron charge at $\nu$=5/2 quantum Hall state". Nature \textbf{452}, 829-834 (2008).
\item Bid, A., Ofek, N., Heiblum, M., Umansky, V. \& Mahalu, D. "Shot noise and charge at the 2/3 composite fractional quantum Hall state". Phys. Rev. Lett. \textbf{103}, 236802 (2009).
\item Chung, Y., Heiblum, M. \& Umansky, V. "Scattering of bunched fractionally charged
quasiparticles". Phys. Rev. Lett. \textbf{91}, 216804 (2003).
\item Milovanovic, M. \& Reed, N. "Edge excitations of paired fractional quantum Hall states". Phys. Rev. B \textbf{53}, 13559-13582 (1996).
\item Lee, S-S., Ryu, S., Nayak, C. \& Fisher, M. P. A. "Particle-hole symmetry at the v=5/2
quantum Hall state". Phys. Rev. Lett. \textbf{99}, 236807 (2007).
\item Overbosch, B. J. \& Wen, X. G. "Phase transition on the edge of the Pfaffian and anti-Pfaffian quantum Hall state". Cond-Mat 0804/0804.2087v1 (2008)
\item Umansky, V., Heiblum, M., Levinson, Y., Smet, Y. Nuebler, J. \& Dolev, M. "MBE growth
of ultra low disorder 2DEG with mobility exceeding 35.106cm2/V-s". J. Crystal Growth \textbf{311},
1658-1662 (2009).
\item M. Dolev, Y. Gross, Y. C. Chung, M. Heiblum, V. Umansky, \& D. Mahalu, "Unexpectedly
large quasiparticles charge in the fractional quantum Hall effect". Phys. Rev. B. \textbf{81}, 161303(R)
(2010).
\end{enumerate}}

\end{document}